\documentstyle[12pt]{article}

\makeatletter                    %allow @ in command names
\@addtoreset{equation}{section}  %reset equation count in each section
\makeatother                     %disallow @ in command names

\begin{document}

\title{Cosmology as `Condensed Matter' Physics
\thanks{Invited talk given at the Third Asia-Pacific Physics Conference,
June 1988, Hong Kong. Published in the Proceedings edited by
Y. W. Chan, A. F. Keung, C. N. Yang and K. Young
(World Scientific Publishing Co, Singapore, 1988) Vol. 1, pp.301-314}}
\author{B. L. Hu
\thanks{Work supported in part by the National Science Foundation under
Grant No. PHY87-17155.}\\
{\small Department of Physics, University of Maryland,
College Park, MD 20742, USA }}
\date{\small {\sl (UMDPP\#89-013, July, 1988)}}
\maketitle
\begin{abstract}
We note that in general there exist two basic aspects
in any branch of physics, including cosmology - one
dealing with the attributes of basic constituents and
forces of nature, the other dealing with how structures
arise from them and how they evolve.  Current research
in quantum and superstring cosmology is directed mainly
towards the first aspect, even though a viable theory
of the underlying interactions is lacking.  We call the
attention to the development of the second aspect,
i.e., on the organization and processing of the basic
constituents of matter (in classical cosmology) and
spacetime (in quantum cosmology).  Many newly developed
concepts and techniques in condensed matter physics
stemming from the investigation of disordered,
dynamical and complex systems can guide us in asking
the right questions and formulating new solutions to
existing and developing cosmological issues, thereby
broadening our view of the universe both in its
formative and present state.
\end{abstract}

\newpage

In this general talk I would like to share with you some
recent thoughts of mine on the direction of cosmological research.
They are based on my observation of the evolution of cosmological
theories in the past two decades$^{1)}$ and my partaking of the research of
quantum processes in the early universe$^{2)}$.  Development of theories
of the early universe$^{3)}$ in the seventies and eighties is partly an
extension of relativistic cosmology developed in the sixties based
on classical general relativity theory.  Its most notable examples
are the so-called "standard model" constructed from the
Friedmann-Robertson-Walker universes and the "chaotic cosmology"
based on the Bianchi universes.  Indeed one major component, the
Kaluza-Klein cosmology, is a class of higher-dimensional anisotropic
universe.  There is also infusion of new ideas from particle physics
and quantum field theory - most notably the inflationary cosmology,
the semi-classical cosmologies, and quantum cosmology.  These new
cosmological theories rely in various degrees on the working of many
gravitational, quantum and statistical processes in the early
universe.  These physical processes have close analogies in condensed
matter physics.  Examples are phase transition in inflationary
cosmology, particle production and backreaction in semiclassical
cosmology, quantum tunneling in quantum cosmology, etc.$^{2-4)}$.  It is
against this backdrop that I shall attempt to make some synthesis of
ideas and introduce some new ones.  I should point out that none of
the specific points discussed here is new, but by rendering them in
some particular ways I hope that new insight will emerge which may
prove useful in guiding future research in cosmology.
\par
By studying the attributes of these models and the nature of
these physical processes, I come to realize that there are two basic
aspects in the formulation of any cosmological model.  One aspect
involves the basic constituents and forces, the other involves the
structure and dynamics, i.e., the organization and processing of these
constituents as mediated by the basic forces or their derivatives.
The first aspect is provided by the basic theories describing
spacetime and matter.  The second aspect in addressing the universe
and its constituents {\sl is} cosmology proper.

\section{Two Basic Aspects in Physics}

It is not difficult to recognize that actually these two
aspects permeate throughout almost all subfields in physics, or
science in general.  Examples of the first aspect in physics dealing
with the "basic" constituents and forces are general relativity,
quantum mechanics, quantum electrodynamics, quantum chromodynamics,
grand unifield theories, supersymmetry, supergravity, quantum
gravity and superstring theories.  The second aspect dealing with
the structure and dynamics is the subject matter of biology,
chemistry, molecular physics, atomic physics, nuclear physics and
particle physics. The former aspect is treated today primarily in
the disciplines of elementary particle physics and general
relativity.  The latter aspect is treated today in the discipline
known collectively as condensed matter physics.  In this sense we
can, for example, regard nuclear physics as condensed matter physcis
of quarks and gluons, the collective manifestation of QCD force.
\par
Note, however, the {\sl duality} and the {\sl interplay} of these two
aspects in any discipline.  On the one hand the basic laws of nature
are often discovered or induced from close examination of the
structure and properties of particular systems - witness the role
played by atomic spectroscopy and scattering in the discovery of
quantum mechanics and atomic theory, accelerator experiments in
advancing particle physics.  On the other hand once the nature of
the fundamental forces and constituents are known one attempts to
depict reality by deducing possible structures and dynamics from
these basic laws.  Thus the study of electrons and atoms via
electromagnetic interaction has been the underlying theme of condensed
matter physics.  Deducing nuclear force from quarks and gluons via QCD
remains the central task of nuclear research today.  From general
relativity one attempts to deduce the properties of neutron stars,
black holes and the universe, which is the theme of relativistic
astrophysics and cosmology.
\par
Note that many known physical forces are not fundamental (in
the sense that they are irreducible), but are effective in nature.
Molecular forces and nuclear forces are such examples.  One may also
regard gravity as an effective force.  Note also that many
disciplines contain dual aspects.  This is especially true in the
developing areas, in which the basic forces and constituents of the
system are not fully understood.  For example particle physics deals
both with the structure and the interactions.  This is seen in the
dialectic relation of, say, quantum flavor and color dynamics and
the duality of compositeness and elementarity.  Similar aspects
exist in superstring and quantum gravity theories.
\par

\section{Two Basic Aspects in Cosmology}

What about cosmology?  The above-mentioned dual aspects are
certainly apparent.  What is new is that in addition to matter (as
described by particles and fields) we have to add in the consideration
of spacetime (as described by geometry and topology).  In the first
aspect concerning constituents and forces, there are also two
contrasting views$^{5)}$:  The "idealist" takes the view that spacetime is
the basic entity, the laws of the universe is governed by the dynamics
of geometry.  Matter is viewed as perturbations of spacetime, particle
as geometrodynamic excitons$^{6)}$.  These ideas, as extention of
Einstein's theory, are not that strange as they may appear.  For
example it is well-known in gravitational perturbation theory that
high frequency perturbations off a background spacetime behave like
relativistic fluid.  Closer to this idea are the Kaluza-Klein and
superstring theories.  There, particles are representations of
internal symmetries, graviton the resonant modes of strings.  Although
a geometric theory is difficult to construct, the philosophical
overtone of these theories is clear.  By contrast the "materalist"
takes the view that spacetime is the manifestation of collective,
large scale interaction of matter fields.  Thus according to
Sakharov$^{7)}$, gravity should be treated as an effective theory, like
elasticity to atomic forces.  This is expressed in the induced gravity
program$^{8)}$.  Despite its many technical difficulties, this view still
evokes some sombering thoughts.  It suggests among others that the
attempt to deduce a quantum theory of gravity by quantizing the metric
may prove to be as meaningful as deducing QED from quantizing
elasticity.  In recent years the apparent contrast between
particle-fields and geometry-topology has dissolved somewhat in the
wake of superstring theory$^{9)}$.  The fact that the same concept can be
viewed in both ways may indeed offer some new insights into the nature
of our universe.
\par
As for the second aspect in cosmology, i.e., the
manifestation of basic forces in astrophysical and cosmological
processes, one sees that almost with any subdiscipline of physics
there is a corresponding branch of astrophysics.  Hydrodynamic and
radiative processes in classical astronomy and astrophysics are
based on Newtonian gravity and Maxwellian electromagnetism.  The
application of nuclear physics to astrophysical phenomena in the
$50's$ and $60's$ has successfully explained nucleosynthesis and neutron
star structure.  The $70's$ and $80's$ saw the advent of particle
astrophysics and inflationary cosmology based on the grand unified
theories, and semiclassical cosmologies based on curved-space
quantum field theory$^{1)}$.  However, the central theme of cosmology
which addresses the state of the universe as a whole is more than
the sum-total of its individual components, as depicted by the many
subdisciplines of astrophysics.  There are broader issues special to
the overall problem of how the universe comes into being and why it
should be the way it is, which touch on the basic problems of
quantum mechanics and relativity theory.  The inquiry of these
issues will necessarily force us back to the first aspect of
cosmological research discussed above.
\par

\section{Two Directions of Cosmological Research}

Depending on the relative emphasis one puts on these two
aspects, current research on cosmological theories follow roughly two
directions:
\par
A)  Cosmology as consequences of quantum gravity and superstring theories.
\par
B)  Cosmology describing the structure and dynamics of the universe
\par
In the first direction, quantum cosmology as represented
historically by the work of Wheeler, DeWitt, Misner, Hawking,
Hartle, Gell-Mann, Coleman and others$^{10)}$ deals more with the
boundary conditions and constraints of quantum gravity as it applies
to our universe than with the theory itself.  Many of the questions
raised in current research such as the wave function, the density
matrix, the vacuum state, the nature of time, conditional
probability, etc., touch on the fundamental problems (especially
their problematic intersections) of general relativity and quantum
mechanics, as manifested in a rather special and unique system which
is our universe$^{11)}$.  Oftentimes we have to extend our consideration
of physics to superspaces$^{12)}$ and other universes$^{13)}$.  In this
context the universe is regarded as a special medium  where the
conflicts of quantum mechanics and relativity are acted out.  It is
not exactly the study of the consequences of a quantum gravity
theory in the same sense as particle astrophysics or inflationary
cosmology with respect to GUT theories.  And it is with this
emphasis that its importance should be properly attached.  Likewise,
many current studies of so-called "superstring cosmology" based on
the picture of spacetime as a smooth manifold is, in my opinion, at
best irrelevant and likely totally wrong.  They are incorrect not
just because they attempt to draw implications without an
established theory, but more so because they do not address the
correct problems.  One exciting aspect of superstring theory is that
not only does it depict a new picture of spacetime based on extended
geometric objects but it provides one with the methodology to quantify
new concepts such as topology change, etc.$^{14)}$.  To say something new,
i.e., different from conventional cosmology based on manifold
spacetime superstring cosmology should at least begin with a different
concept of spacetime.  Doing it otherwise misses the whole point.
\par
In this first direction one could also include inquiries or
proposals made which view the universe as manifestor of physical laws,
as formulator of rules, as processor of information, etc.$^{15)}$.  This
direction of cosmological research touches on the basic laws of
quantum mechanics, general relativity and statistical mechanics.  In
this field the formulation of meaningful problems are almost as
important as seeking their solutions.  Progress will be slow but the
intellectual reward is profound.
\par

\section{Cosmology as `Condensed Matter' Physics}

By now I hope the meaning of this figurative description is
clear.  Please bear in mind that by "condensed state" I refer both
to matter and spacetime.  Cosmology is the study of the organization
and processing of matter as well as spacetime points.  To be
explicit I have sketeched in Table I some major ingredients of
condensed matter physics, nuclear physics and the physics of the
early universe.  The early universe is included here because it
invokes many physical processes which directly affect the overall
structure and dynamics of spacetime (e.g. particle production and
backreaction, quantum vacuum energy and inflation, etc.).  Table II
outlines the major themes of recent development of condensed matter
physics.  Notice the increasing importance attached to nonlinear,
nonlocal and stochastic behavior of complex systems.  The neighboring
column lists problems of a similar nature in the cosmology of the
early universe.  In contrast to Table I, the problems listed here are
mainly representative in nature and are largely undeveloped.  Such a
comparison is aimed at stimulating new thoughts along these lines.
None of these ideas are due exclusively to me, nor are they completely
new - many of them have been toyed with some twenty years ago$^{5,6)}$.
The difference between now and then is that 1) concepts and techniques
in particle physics, especially superstring theory, have developed to
the degree that the mathematical formulation of these problems has
become possible, and 2) advances in condensed matter physics such as
phase transition and critical dynamics, order-disorder behavior,
dynamical systems, complex systems, etc. have opened up new
possibilities in probing the organization and dynamics of matter in
various states.  These techniques and ideas may provide useful hints
in understanding how spacetime takes shape, how the universe evolves,
what determines its content and how its many different structural
forms develop.
\par
For illustrative purpose I have listed in Table III some
sample problems of this nature in cosmology, with respect to the
universe in its present, early and primeval states.  I have only
mentioned the essential underlying ideas, with some sample references,
should the reader be interested in the details.
\par
By organizing these problems according to some general theme
and by providing some overall perspective, novel as it may be, I
hope this could generate some interest in pursuing cosmological
research in a new light - as "condensed matter" physics of general
relativity and quantum gravity.
\par

\section{New Elements}

In examining these new problems and concepts we see that
according to this view two major ingredients will likely contribute
to shaping a new direction of cosmological research:  One is
topology and the other is stochasticity, both  for matter-field and
spacetime-geometry systems.  For the concepts of spacetime-geometry
what is more important is not geometry, but topology; not topology,
but point sets.  For basic laws, one's focus moves from the rules to
construct content to the rules to construct rules.  As for
structure, what is more important is not regularity, but
randomness: not order, but chaos - or, more interestingly, order
out of chaos; not simplicity, but complexity - or complexity out of
simplicity.  In summary, the new direction seems to be forged with
topological ideas applied to spacetime and fields and statistical
ideas applied to structures and basic laws.  Cosmological research
would benefit from recognizing and harnessing these new
developments.

\newpage

\centerline{TABLE I - BRANCHES OF "CONDENSED MATTER" PHYSICS}
\vskip 1cm
\begin{tabular}{lll}
CONDENSED MATTER PHYSICS &   NUCLEI AND PARTICLES &    EARLY UNIVERSE \\
\hline
&&\\
- electromagnetic interaction & - strong interaction  &    - quantum fields \\
                              &                       &   ~~  in curved space
\\
&&\\
{\sl CONSTITUENTS} \\
& & \\
- electrons       &        - quarks, gluons    &      - particle-fields \\
- atoms           &        - mesons, baryons   &     - spacetime-geometry  \\
&&\\
{\sl FORCES}    \\
&&\\
- electronic-ionic   &        - chromo-    &         - general relativity \\
                      &    ~~electromagnetism  &     ~~    + GUT   \\
- chemical-     &        - nucleon force     &       - gravity as effective \\
{}~~ molecular bonds                       &           & ~~ force\\
&&\\
{\sl COLLECTIVE EXCITATIONS} \\
&&\\
- lattice and electron  &    - particle spectrum  & - Casimir effect and \\
- phonon        &          - bound states   &         particle creation from\\
- plasmon        &         - resonances    &           magnification of   \\
- exciton         &        - solitons, skyrmions  &   quantum fluctuations   \\
                  &                             & - graviton and particles as
\\
               &                         & excitation of spacetime \\
&&\\
{\sl PHASE TRANSITIONS} \\
&&\\
- solid-liquid-gas &   - quark-hadron      & - of Higgs field at $t_{GUT}$:\\
- superconductivity &  ~~ phase transition & ~~  inflationary  transition \\

- metal-insulator, etc.  &             & - of spacetime at $t_{Planck}:$\\
                &                      & black hole-string transition\\
\end{tabular}

\newpage
\centerline{TABLE II - DEVELOPMENT OF "CONDENSED MATTER" PHYSICS}
\vskip 1cm
\begin{tabular}{ll}
CONDENSED MATTER PHYSICS  &                         COSMOLOGY \\
\hline
{\it as the organization and processing of} \\
&\\
atoms and electrons     &                        spacetime and matter \\

&\\
{\sl FRAMEWORK OF SYSTEMS:} & {\sl NONLINEARITY, NONLOCALITY, STOCHASTICITY}\\
&\\
l.  {\sl Ordered Systems}\\
&\\
lattice + electrons    &     - spacetime (as smooth manifolds)+ perturbations\\

+ excitations          &     - fields + fluctuations  \\

&\\
2.  {\sl Disordered System}\\

&\\

- topological defects:  &        - topological structures of gauge fields \\
{}~~strings, domain walls, etc   &      - multiply-connected spacetimes  \\

&\\
3.  {\sl Random System}\\
&\\
amorphous state       &        - random fields in curved spacetime  \\

spin-glass            &        - stochastic spacetimes            \\

random network        &                                          \\

&\\
4.  {\sl Dynamical System}\\
&\\
- chaos-order         &        {\sl  organization and processing of}   \\

- metric and topological   &   - matter (e.g. galaxy distribution)  \\
{}~~entropy                    &                                         \\

- fractals                 &   - spacetime (e.g. chaos in mixmaster)  \\

&\\
5.  {\sl Complex System}\\
&\\
- spin glass            &      - self-reproducing universes   \\

- neural network,       &      - spacetime as organization of  \\
{}~~ parallel processor      &   ~~ points (e.g. Borel sets)$^{15}$  \\

- information theory    &      - physical laws as processing of   \\

                       & ~~  propositons (e.g. quantum logic)$^{15}$  \\

\end{tabular}

\newpage

\centerline{TABLE III - SOME SAMPLE PROBLEMS}
\vskip 1cm
\begin{tabular}{lll}
  PROBLEMS          &                THEMES    &         SAMPLE\\
\hline
                    &                          &    REFERENCES \\
&&\\

A. {\sl Present Universe}&&\\

&&\\
l.  galaxy correlation function  &      fractal dimension   &     16  \\

2.    voids and foam-like structure  &  topology of matter distribution & 17 \\

3.   galaxy formation         &     cosmic strings, topology &       18  \\
                              &  of field  configurations    &   \\

&&\\
B. {\sl Early Universe}\\
&&   \\

l.  chaos in Bianchi cosmology   &      chaotic dynamics -  &     19  \\
                                      &  topological entropy &  \\

2.  strange attractor in   Kaluza-Klein     &  dynamical systems &     20    \\
{}~~~ and superstring cosmology  & &   \\

3.  "self-reproducing" universes  &   cellula automata   &      21,22  \\

4.   hierarchical universes         &    complex systems   &       23,24  \\

&&\\
C.  {\sl Primordial Universe} \\
&&\\
l.   Regge calculus, lattice universe &  - simplicial complex  &   25  \\

2.  spacetime foam        &             - topological classes, &  14   \\
                          &    probability distribution &  \\

3.   stochastic fields and spacetimes &  - stochastic calculus &   26  \\

4.   "Birth" of the universe      &      as interface dynamics &   27,28   \\

5.   causal structure     &            from ultrametricity  &    29,30  \\

\end{tabular}

\newpage
\centerline{REFERENCES}
\vskip 1cm
\noindent
1.   See, e.g. Hu, B. L., "Recent Development of Cosmological
Theories" in G. E. Tauber and B. L. Hu (ed) {\sl Cosmology} - {\sl A Note
and Source Book} (Riedel Publishing Co., Dordrecht, 1989).
\\

\noindent
2.   See, e.g., Hu, B. L., in Proceedings of the 2nd (1979) and 4th
(1985) Marcel Grossmann Meetings edited by R. Ruffini (North
Holland Publishing Co. 1982, 1986) and references therein.
\\

\noindent
3.   See, e.g., Hu, B. L., "Quantum Theories of the Early Universe -
A Critical Appraisal" in Proceedings of the International
Conference on Gravitation and Cosmology (Dec. 1987, Goa, India)
edited by C. V. Vishveshwara (Cambridge University Press 1989)
and references therein. \\

\noindent
4.   See, e.g. Hu, B. L., "On the Nature of Quantum Processes in the
Early Universe" in Proceedings of the 5th Marcel Grossmann
Meeting, August 1988, Perth, Australia. \\

\noindent
5.   See, e.g, Misner, C. W., Thorne, K. S., and Wheeler, J. A.,
{\sl Gravitation} (Freeman, San Francisco, 1973) Sec. 44.4. \\

\noindent
6.   See, e.g., Wheeler, J. A. {\sl Geometrodynamics} (Academic Press, N.Y.
1962); and private discussion in the Centenary Symposium in Honor
of Clifford's 1871 paper (Princeton University 1971).\\

\noindent
7.   Sakharov, A. D., Doklady Akad. Nauk. S.S.S.R. 177 (1967) 70. \\

\noindent
8.   Adler, S. L., Rev. Mod. Phys. {\sl 54}, 729 (1982). \\

\noindent
9.   See, e.g., {\sl Superstring Theories}, Vol. $1 \& 2$ by M. Green, J.
Schwarz and E. Witten (Cambridge Univ. Press 1987).\\

\noindent
10.  Wheeler, J. A., in {\sl Battelle Rencontres} ed. by C. M. DeWitt and
J. A. Wheeler (Benjamin, New York 1968).  DeWitt, B. S., Phys. Rev.
{\sl 160,} 1113 (1967).  Misner, C. W., in {\sl Magic Without Magic} (ed. by
J. Klauder (Freeman, San Francisco 1970).  Hartle, J. B. and
Hawking, S. W., Phys. Rev. {\sl D28}, 2960 (1983).
Vilenkin, A., Phys. Lett. {\sl 117B}, 25 (1982); Gell-Mann, M., in
Fermilab Workshop on Quantum Cosmology (1987); Coleman, S.,
Harvard University preprint (1988).
For recent reviews and references on quantum cosmology, see,
e.g., J. B. Hartle in the Yale Advanced Study Institute (1985);
S. W. Hawking in Physica Scripta (1986).\\

\noindent
11.  Gell-Mann, M., "Quantum Mechanics and Our Specific Universe"
(1987).\\

\noindent
12.  Ch. 43 of Ref. 5.\\

\noindent
13.  Everett, H., Rev. Mod. Phys. {\sl 29}, 454 (1957); {\sl The Many-Worlds
Interpretation of Quantum Mechanics} ed. by B. S. DeWitt and N.
Graham (Princeton University Press 1973).\\

\noindent
14.  See, e.g. J. A. Wheeler "Geometrodynamics and the Issue of the
Final State" in {\sl Relativity, Groups and Topology} ed. by B.
DeWitt and C. DeWitt (Gordon and Breach, 1964); S. W. Hawking
in {\sl General Relativity} - {\sl An Einstein Centenary Volume} ed. by S. W.
Hawking and W. Israel (Cambridge University Press, 1979).  A.
Anderson and B. S. DeWitt (1987); A. Anderson, "Changing
Toplogy and Non-Trivial Homotopy" Maryland preprint (1988); and
recent work by E. Witten and co-workers on superstring theory. \\

\noindent
15.  See, e.g. Chap. 44 of Ref. 5, See also Wheeler, J. A.,
"From Relativity to Mutability" in {\sl The Physicist's Conception of
Nature} ed. J. Mehra (Reidel, Dordrecht 1973)
"Frontiers of Time" in {\sl Problems in the Foundations of Physics}
Enrico Fermi LXXII Course, Varenna, ed. N. Toraldo di
Francia and Bas van Fraassen (N. Holland, Amsterdam 1979)
"The Austerity Principle in Physics"; China lectures (1982);
"Observer-Participancy and Quantum Physics" talks (1983-87).\\

\noindent
16.  Szalay, A. S. and Schramm, D. N., Nature {\sl 314}, 718 (1985). \\
Bahcall, N. and Burgett, W. S., Ap. J. {\sl 300}, L35 (1986). \\
Mandelbrot, B., {\sl Fractals} (W. H. Freeman 1977). \\

\noindent
17.  de Lapporent, V., {\sl et al}, Ap. J. Lett. {\sl 302,} Ll (1986).  For
recent work see A. Mellot (ed.) Proc. Conference on Topology and
Large Scale Structure of the Universe (1988). \\

\noindent
18.  Kibble, T.W.B., Phys. Rep. {\sl 67}, 183 (1980); Vilenkin, A., ibid
{\sl 121,} 263 (1985); Brandenburger, R. and Turok, N. in Proceedings
of the Goa Meeting, C. V. Vishveshwara (ed.) (Cambridge
University Press 1988).\\

\noindent
19.  Barrow, J. D., Phys. Rep. {\sl 85}, l (1983).\\

\noindent
20.  Maeda, K., Class. Quant. Grav. {\sl 3}, 233, 651 (1986); Phys. Rev.
{\sl D35}, 471 (1987).\\

\noindent
21.  Linde, A. D., Mod. Phys. Lett. {\sl A1,} 81 (1986); Phys. Lett. {\sl l75B}
395 (1986). \\

\noindent
22.  Wolfram, S., {\sl Cellula Automata} (World Scientific 1987).\\

\noindent
23.  For a simple introduction, see, e.g., R. Serra, {\sl et al
Introduction to Complex Systems} (Pergamon Press, Oxford 1986).
For recent research, see, e.g., {\sl Complex Systems}, journal edited
by S. Wolfram.\\

\noindent
24.  See, e.g., Palmer, R. G., Adv. Phys. {\sl 31}, 669 (1982) and Ref.
30. Hu, B. L., "Hierarchical Universe Models from Effective
Energy Terrains" (unpublished).\\

\noindent
25.  Chap. 42 of Ref. 5; H. Hamber and R. William, 1{\sl 57B}, 368 (1985);
J. B. Hartle, J. Math. Phys. {\sl 26}, 804 (1985); H. Hamber,
"Simplicial Quantum Gravity" in {\sl Critical Phenomena, Random
Systems, Gauge Theories}, ed. by K. Osterwalder and R. Stora (N.
Holland, Amsterdam 1986).\\

\noindent
26.  See, e.g., Hertz, J., {\sl Disordered Systems}, Physica Scripta {\sl T10},
 l (1985), Thouless, D. J., in 1984 Les Houches Summer School (Ref.
25); Mazard, M., {\sl et al, Spin Glass Theory and Beyond} (World
Scientific, Singapore, 1987).\\

\noindent
27.  Vilenkin, A., in Ref. 10; Sato, K. {\sl et al}, Phys. Lett {\sl 108B}, 103
(1982); Blau, K., Guth, A., and Guendelman, E. I., {\sl D35,} 1747
(1985); Giddings, S. and Strominger, A., Princeton IAS preprint
(1987).\\

\noindent
28.  See, e.g., {\sl Phase Transition and Critical Phenomena}, Vol. 11, ed.
by C. Domb and J. L. Lebowitz (Academic Press, 1987).
Hu, B. L., "Birth of the Universe as Interface Dynamics"
(unpublished).\\

\noindent
29.  See, e.g., t'Hooft, G., in 8th International Conference on
{\sl General Relativity and Gravitation}, Waterloo, Canada 1977;
Sorkin, R., Princeton IAS preprint (1987). \\

\noindent
30.  For related concepts, see, e.g., Rammal, R., Toulouse, G. and
Virasoro, M., Rev. Mod. Phys. {\sl 58,} 765 (1986).

\end{document}